%% file: main.tex
\documentclass[runningheads]{llncs}

\usepackage{iciap}

\usepackage{iciapabbrv}
\usepackage{booktabs}
\usepackage{tabularx}
\usepackage{array}
\usepackage{siunitx}

\usepackage[table]{xcolor}   
\usepackage{wrapfig}

\usepackage{graphicx}
\usepackage{amsmath}
\usepackage{amssymb}
\usepackage{booktabs}
\usepackage{array}
\usepackage{multirow}
\usepackage{color}
\usepackage{colortbl}
\usepackage{framed}
\usepackage{bm}
\usepackage{bbm}
\usepackage{xspace}
\usepackage{enumitem}
\usepackage{cite}
\usepackage{xparse}
\usepackage{xcolor}
\usepackage{algorithm}
\usepackage{lipsum}
\usepackage{listings}
\usepackage{wrapfig}
\usepackage{adjustbox}
\usepackage[table]{xcolor}
\usepackage{makecell} 
\usepackage[accsupp]{axessibility}  

\definecolor{blond}{RGB}{250, 240, 190} 

\usepackage{hyperref}

\usepackage{orcidlink}

\newcommand{\clip}{CLIP\cite{radford2021learning}}
\newcommand{\mae}{MAE\cite{he2022masked}}
\newcommand{\siglip}{SigLIP\cite{zhai2023sigmoid}}
\newcommand{\dinovd}{DINOv2\cite{oquab2024dinov2}} 

\newcommand{\clipfull}{\textit{CLIP-vit-base-patch32} \cite{radford2021learning}}
\newcommand{\maefull}{\textit{MAE-vit-base}\cite{he2022masked}}
\newcommand{\siglipfull}{\textit{SigLIP-base-patch16-224}\cite{zhai2023sigmoid}}
\newcommand{\dinovdfull}{\textit{DINOv2-base}\cite{oquab2024dinov2}} 

\newcommand{\deepseeks}{DeepSeek-R1-8B\cite{guo2025deepseek}}
\newcommand{\deepseekb}{DeepSeek-R1-70B\cite{guo2025deepseek}}
\newcommand{\gemma}{Gemma2-9B\cite{gemma_2024}}
\newcommand{\chatgpt}{ChatGPT-4o\cite{achiam2023gpt}}
\newcommand{\llamas}{Llama 3.1 8B\cite{grattafiori2024llama}}
\newcommand{\llamab}{Llama 3.1 70B\cite{grattafiori2024llama}}
\newcommand{\mistralb}{Mistral 24B}
\newcommand{\mistrals}{Ministral 8B}

\newcommand{\clipscore}{CLIPScore\cite{hessel2021clipscore}}
\newcommand{\dinosimilarity}{DINO similarity\cite{oquab2024dinov2}}
\newcommand{\lpips}{LPIPS\cite{zhang2018perceptual}}
\newcommand{\fid}{FID\cite{NIPS2017_8a1d6947}}

\newcommand{\blip}{Blip2\cite{10.5555/3618408.3619222}}
\newcommand{\idefics}{Idefics3 \cite{laurençon2024building}}
\newcommand{\florence}{Florence2 \cite{xiao2023florence}}

\newcommand{\tit}[1]{\smallbreak\noindent\textbf{#1.}}

\def \ie {\emph{i.e.}}
\def \eg {\emph{e.g.}}

\newcommand{\ours}{{SVGauge}\xspace}
\newcommand{\dataset}{SHE\xspace}

\begin{document}

\title{\ours: Towards Human-Aligned\\ Evaluation for SVG Generation} 


\author{Leonardo Zini\inst{1}\orcidlink{0009-0003-9439-9867}, 
Elia Frigieri\inst{1} \orcidlink{0009-0005-5310-5169}, 
Sebastiano Aloscari\inst{1}\orcidlink{0009-0008-3588-6848},  \\
Marcello Generali\inst{2}, 
Lorenzo Dodi\inst{2}, 
Robert Dosen \inst{2}, 
Lorenzo Baraldi\inst{1}\orcidlink{0000-0001-5125-4957}}

\authorrunning{L. Zini et al.}

\institute{
University of Modena and Reggio Emilia 
\and
Doxee S.p.A.
}

\maketitle

\begin{abstract} Generated Scalable Vector Graphics (SVG) images demand evaluation criteria tuned to their symbolic and vectorial nature: criteria that existing metrics such as FID, LPIPS, or CLIPScore fail to satisfy. In this paper, we introduce SVGauge, the first human-aligned, reference-based metric for text-to-SVG generation. SVGauge jointly measures (\textit{i}) visual fidelity, obtained by extracting SigLIP image embeddings and refining them with PCA and whitening for domain alignment, and (\textit{ii}) semantic consistency, captured by comparing BLIP-2-generated captions of the SVGs against the original prompts in the combined space of SBERT and TF-IDF. Evaluation on the proposed SHE benchmark shows that SVGauge attains the highest correlation with human judgments and reproduces system-level rankings of eight zero-shot LLM-based generators more faithfully than existing metrics. Our results highlight the necessity of vector-specific evaluation and provide a practical tool for benchmarking future text-to-SVG generation models.
\end{abstract}

\section{Introduction}
\label{sec:intro}
\input{sections/introduction}

\section{Related Works}
\label{sec:related}
\input{sections/related}

\section{Proposed approach}
\label{sec:method}
\input{sections/method}

\section{Experimental Evaluation}
\label{sec:experiments}
\input{sections/experiments}

\section{Conclusion}
\input{sections/conclusion}

\section{Acknowledgments}
We acknowledge the CINECA award under the ISCRA initiative, for the availability of high-performance computing resources. This work has been conducted with the support of the PRIN 2022-PNRR project "MUCES" (CUP E53D23016 290001), the PRIN 2022 project "MUSMA"' (CUP E53D23008310001) and the European project MINERVA, funded by European High-Performance Computing Joint Undertaking (JU) under grant agreement No 101182737.

%
%
\bibliographystyle{splncs04}
\bibliography{main}
\end{document}

%% file: sections/introduction.tex
The objective of text-to-SVG generation is to produce vectorial graphical representations conditioned on natural language prompts, accurately conveying the intended semantics while respecting the stylistic and structural properties of SVGs. Unlike raster images, SVGs offer a symbolic, abstract, and resolution-independent representation of visual concepts, often emphasizing geometric structures and minimalistic design choices. As such, the task demands not only understanding the textual description but also capturing fine-grained visual abstractions and compositional relationships between elements.

Recent advances in generative modeling have considerably improved the quality of SVG synthesis, with innovative strategies including rasterization-then-vectorization pipelines~\cite{ma2022towards}, differentiable vector renderers~\cite{li2020differentiable}, and latent diffusion frameworks adapted to the vector space~\cite{jain2023vectorfusion,xing2024svgdreamer,xing2025svgfusionscalabletexttosvggeneration}. Additionally, there has been a growing interest in leveraging Large Language Models (LLMs) for auto-regressive SVG token generation~\cite{CIT_IconShop,xing2024llm4svg,wu2024chat2svg,rodriguez2024starvector}, harnessing the human-readable nature of SVG markup for more interpretable synthesis.

As generation quality continues to progress, the need for effective evaluation becomes more critical. Current evaluation methods largely rely on metrics designed for raster images, such as Fréchet Inception Distance (FID)\cite{heusel2017gans}, Learned Perceptual Image Patch Similarity (LPIPS)\cite{zhang2018perceptual}, or DINOv2-based similarities~\cite{oquab2024dinov2}, or on text-image alignment scores like CLIPScore~\cite{hessel2021clipscore}. However, these approaches often fall short when applied to SVGs: they fail to capture the symbolic, geometric, and stylistic nuances intrinsic to vector graphics, and can exhibit strong sensitivity to low-level changes despite semantic equivalence.

\begin{figure}[t]
     \centering
     \includegraphics[width=0.98\linewidth]{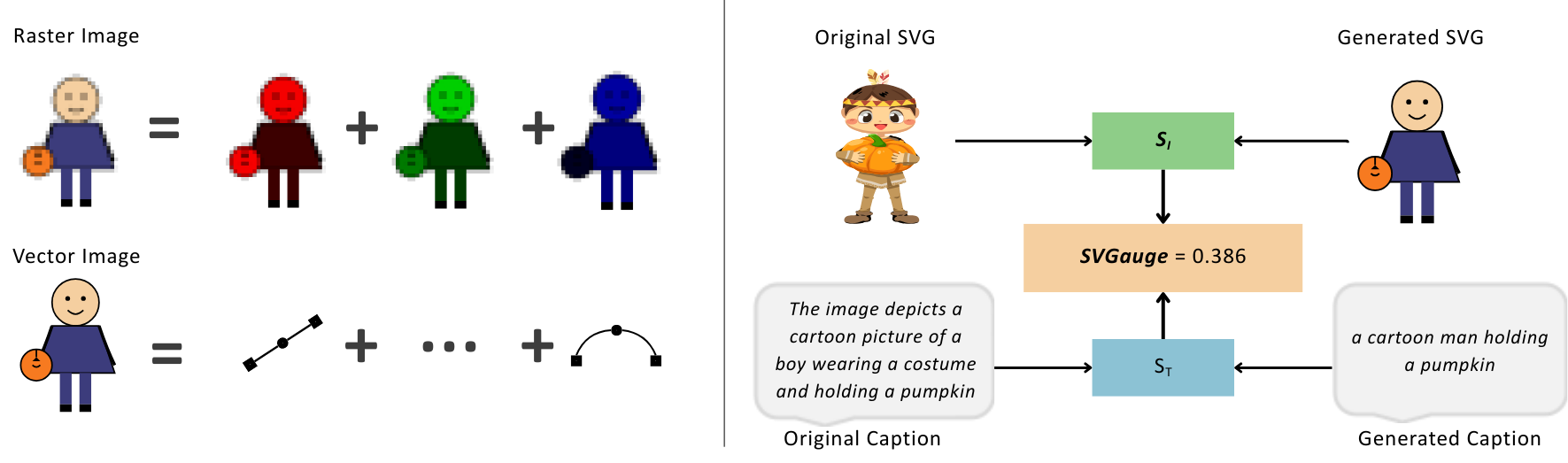}
     \caption{Comparison between raster (\eg, JPG) and vector images (\eg, SVG) and an overview of \ours for SVG generation evaluation.}\label{Fig:first_page}
     \vspace{-1em}
\end{figure}

In light of these limitations, we argue that evaluating text-to-SVG generation requires dedicated metrics that specifically account for the distribution of SVG images and align closely with human perceptual and semantic judgment. In response, we propose \ours, a novel metric for assessing the quality of text-to-SVG generations, designed to robustly capture both visual similarity in a vector-aware embedding space and semantic preservation via multimodal text analysis (Fig.~\ref{Fig:first_page}). Our approach begins by rasterizing both the reference and generated SVGs to enable feature extraction via pre-trained vision backbones, followed by domain adaptation techniques including PCA and whitening. Visual similarity is computed by comparing embeddings in this adapted space, while semantic consistency is evaluated through a captioning loop that uses a multimodal LLM to generate descriptions of the SVGs, which are then compared using a combination of Sentence-BERT embeddings and TF-IDF-weighted similarity.

Extensive experiments demonstrate that \ours achieves superior alignment with human judgment across a wide range of SVG generation scenarios, outperforming conventional raster-based metrics. We show that our dual-axis design provides robustness and increased alignment with human evaluations, better reflecting the qualities valued in SVG synthesis. As a complementary contribution, we also develop and release the first dataset with prompt-SVG pairs annotated with human evaluations.

%% file: sections/related.tex

Here, we provide a concise overview of the most relevant works related to SVG generation using Latent Diffusion Models or LLMs, and quality measurement.

\tit{Latent Diffusion Models}
\label{sec:ldm_gen}
A common approach to SVG generation synthesizes a raster image from a text prompt using latent diffusion models, followed by vectorization via traditional tools like LIVE~\cite{ma2022towards}, VTracer, or Potrace. DiffVG~\cite{li2020differentiable} introduced a differentiable renderer enabling SVG optimization through backpropagation with image-based losses such as Score Distillation Sampling (SDS).
VectorFusion~\cite{jain2023vectorfusion} initializes an SVG with fixed paths and optimizes it using latent SDS gradients, while SVGDreamer~\cite{xing2024svgdreamer} improves initialization using activation maps but at the cost of a complex pipeline.
Recently, SVGFusion~\cite{xing2025svgfusionscalabletexttosvggeneration} proposed a unified architecture based on a Vector-Space Diffusion Transformer and a Vector-Pixel Autoencoder to improve both efficiency and quality.


%

\tit{Large Language Models}
\label{sec:llm_gen}
The structured nature of SVGs has also motivated LLM-based generation approaches. IconShop~\cite{CIT_IconShop} trained a Transformer decoder to produce simple icons from text, while LLM4SVG~\cite{xing2024llm4svg} built a large dataset of 250,000 auto-captioned SVGs and fine-tuned compact LLMs with instruction-tuned prompts.
Chat2SVG~\cite{wu2024chat2svg} developed a prompting pipeline for SVG generation, and StarVector~\cite{rodriguez2024starvector} proposed a multimodal model that translates raster images into SVGs by predicting SVG tokens with a large language model.

\tit{Evaluation Methods for SVG generation}
\label{sec:eval_methods}
Current evaluation methods for text-to-SVG rely on metrics developed for raster images, such as DINOv2-based similarity~\cite{oquab2024dinov2}, FID\cite{heusel2017gans}, and LPIPS\cite{zhang2018perceptual}. Text-image alignment metrics like CLIPScore~\cite{hessel2021clipscore} have also been employed. However, these methods can be unreliable for SVGs: being built on backbones optimized for natural images, they often fail to capture the abstract, symbolic, or geometric semantics typical of SVGs. For instance, an icon with a unique stylistic interpretation might be penalized by CLIPScore even if it faithfully represents the input caption. These limitations result in reduced alignment with human perception, which ultimately hinders the development of generative models, as they are evaluated with metrics that poorly reflect true quality and fail to reward meaningful improvements.

%% file: sections/method.tex
We argue for moving beyond the application of metrics developed for images coming from the raster domain and instead propose a metric specifically designed for the SVG domain, carefully aligned with human judgment. Our approach takes a dual perspective, accounting for both visual similarity and semantic preservation, and demonstrates state-of-the-art alignment with human evaluations.
%

\begin{figure*}[t]
    \centering
    \includegraphics[width=\linewidth]{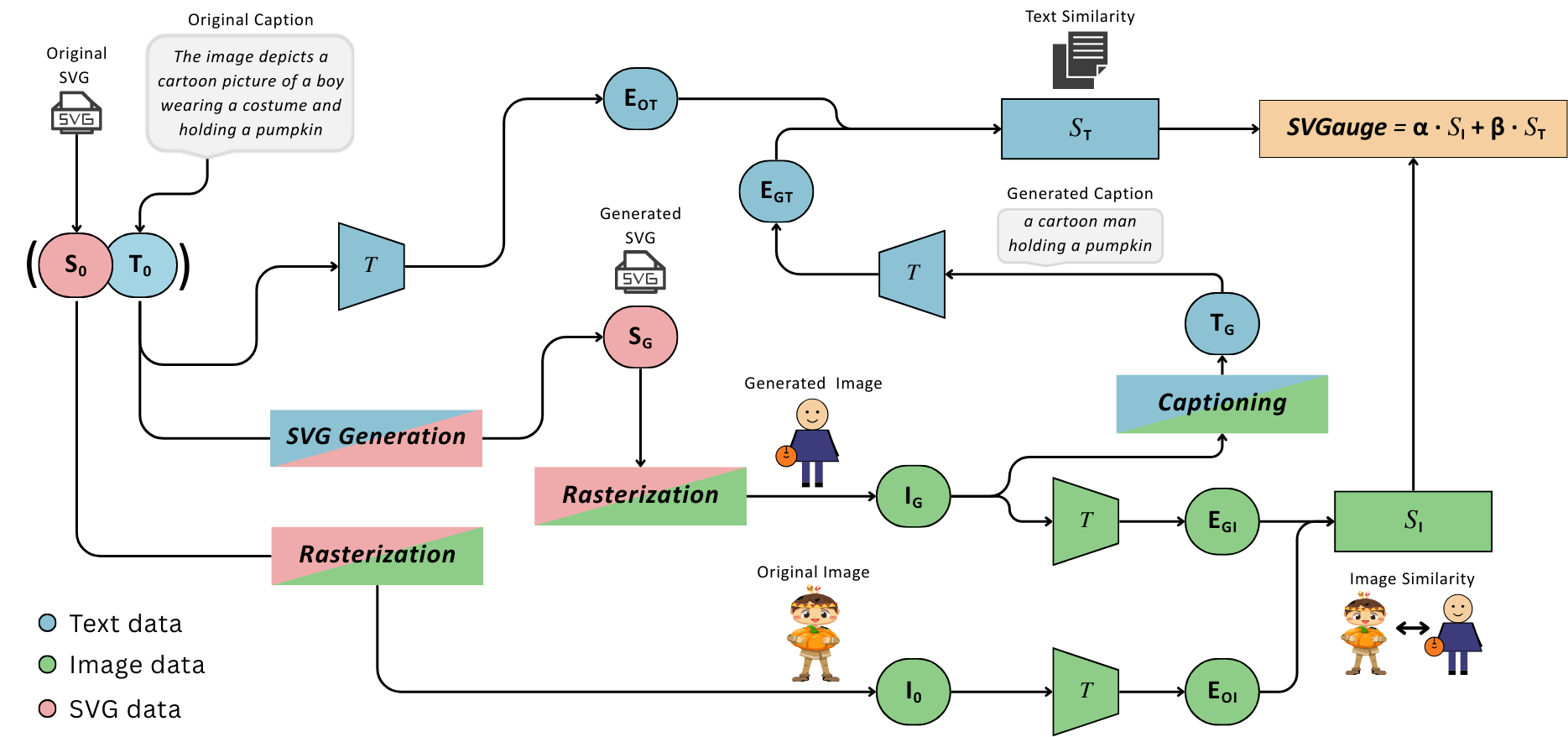}
    \vspace{-.5cm}
    \caption{Overview of the \ours metric for text-to-SVG generation evaluation.
    }
    \label{fig:metric_pipeline}
    \vspace{-1em}
\end{figure*}

\tit{Overview of our approach}
Our approach for text-to-SVG evaluation begins with a reference pair $(T_O, S_O)$, where $T_O$ represents a natural language prompt describing a visual concept, and $S_O$ is an SVG image that accurately depicts this concept. Given a generator $\mathcal{G}$ which takes the prompt $T_O$ as input, we evaluate the quality of its generation, $S_G = \mathcal{G}(T_O)$ through a dual-axis evaluation that captures both the visual resemblance of the generated vectorial image, and the semantic consistency between the reference and generated outputs.

\tit{Rasterization and Image Encoding}
Since SVGs are vectorial representations and not directly compatible with standard image similarity techniques, we first rasterize both the reference $S_O$ and generated $S_G$ into pixel-based images, denoted as $I_O$ and $I_G$, respectively. This rasterization step enables the use of pre-trained vision backbones operating on image tensors.

Each rasterized image is passed through a visual backbone, producing high-dimensional embeddings $E_{OI}$ and $E_{GI}$. To enhance spatial localization, we extract a grid of feature vectors (\eg, from the final self-attention layer of a ViT backbone) and compute their average. Formally, the embedding for the reference image is defined as 
\begin{equation}
    E_{OI} = \frac{1}{H \times W} \sum_{i,j} \text{backbone}(I_O)_{\left[ i, j\right]},
\end{equation}
where $(H, W)$ denotes the feature grid dimensions. The embedding for the generated image is computed analogously.

To further refine these embeddings for similarity computation and adapt them to the domain of vectorial images, we apply a two-stage post-processing pipeline consisting of PCA and whitening, which have been demonstrated to improve retrieval performance in high-dimensional spaces~\cite{jegou:hal-00722622}. Specifically, PCA projects each embedding onto a lower-dimensional subspace by (\textit{i}) centering the data via mean subtraction,
$\mu$, computed over the distribution of vectorial images, \ie,~\[X_{\text{centered}} = X - \mu,\] 
where $\mu$ is the mean vector over a distribution of vector images, and (\textit{ii}) projecting onto the principal components $P$ associated with the largest eigenvalues of the covariance matrix:
\[
X_{\text{PCA}} = P^{\mathsf{T}} X_{\text{centered}}.
\vspace{0.55em}
\]
This projection not only reduces noise but also emphasizes the most discriminative directions, as it ensures that the projected components are decorrelated -- \ie,~for any two different eigenvectors $p_i$ and $p_j$, $p_i^{\mathsf{T}} p_j = 0$. Further, being the projection estimated on a distribution of SVG images, it also reprojects the original embeddings into a subspace which is tailored to the distribution of vectorial images.

Following PCA, we apply a whitening transformation to further normalize the embeddings. Operating in the reduced $d'$-dimensional subspace identified by PCA, whitening rescales each principal component to unit variance, effectively normalizing the covariance matrix to the identity. Let $\lambda_1, \ldots, \lambda_{d'}$ denote the eigenvalues associated with the selected eigenvectors. The whitening transformation is defined as
\vspace{1em}
\[
\hat{X} = \operatorname{diag}\left(\lambda_1^{-\frac{1}{2}}, \ldots, \lambda_{d'}^{-\frac{1}{2}}\right) \, X_{\text{PCA}},
\vspace{1em}
\]
where each component is individually scaled to achieve decorrelation and variance equalization. The resulting embeddings $\hat{X}$ exhibit a balanced distribution, ensuring that no direction dominates the similarity computation and all feature dimensions contribute equally.

\tit{Visual Similarity Evaluation}
Overall, the combination of averaging, PCA, and whitening yields compact and robust image representations, aligning with the principles of negative evidence and decorrelation \cite{jegou:hal-00722622, NIPS2017_8a1d6947}. Finally, we compute the cosine similarity between the embedding of the reference and generated image, to obtain a visual similarity score $S_I$. The aforementioned similarity score quantifies how similar the generated image $I_G$ is to the reference $I_O$ in the learned embedding space \cite{zhang2018perceptual}.

\tit{Semantic Evaluation via Captions}
While visual similarity provides a measure of surface-level resemblance, it may not fully capture whether the generated image semantically reflects the input prompt. An image can be structurally similar to the reference yet fail to convey the intended meaning.

To address this, we introduce a semantic evaluation loop. A multimodal LLM, such as BLIP-2~\cite{10.5555/3618408.3619222}, is employed to generate a caption $T_G$ for the generated image $I_G$, capturing its semantic content. We then compare $T_G$ to the original prompt $T_O$ by encoding both with Sentence-BERT (SBERT)~\cite{reimers2019sentence}, obtaining dense sentence embeddings $E_{GT}$ and $E_{OT}$. Semantic similarity is initially measured via cosine similarity between these embeddings.

However, SBERT may overestimate similarity for short or generic sentences. To mitigate this, we integrate a TF-IDF weighting mechanism that emphasizes rare, informative terms. By doing so, matches on distinctive words are rewarded, while overlaps on common or generic terms are downweighted. The final semantic similarity score is computed as


\begin{equation}
    S_T \;=\;
        \text{CosineSim}\!\left(E_{OT},\,E_{GT}\right)
        \;\cdot\;
        \Bigl(
            0.8 \;+\;
            0.2\,\text{CosineSim} \bigl(V_{OT},\,V_{GT}\bigr)
        \Bigr),
    \label{eq:sim_a_tfidf_cos}
    \vspace{.5em}
\end{equation}
where $E_{OT}$, $E_{GT}$ are the SBERT-space embeddings of the reference and generated texts, respectively, and $V_{OT}$, $V_{GT}$ are their corresponding TF-IDF vectors.
The TF-IDF factor is normalized within the range $[0.8, 1]$ to act as a gentle rescaling term that adjusts similarity based on the content informativeness of the caption match.

\tit{Combined Evaluation Metric}
Our final evaluation metric combines both visual and semantic similarities to form a unified score, defined as
\vspace{.55em}
\[
    \mathrm{\ours} = \alpha \cdot S_I + \beta \cdot S_T.
    \label{eq:final_metric}
\vspace{.5em}
\]

Here, $\alpha$ and $\beta$ are scalar weights that allow us to tune the importance of each component based on the use case. For instance, applications that prioritize aesthetic consistency may favor $\alpha > \beta$, while semantic-critical applications (\ie,~ educational content generation) may prefer $\beta > \alpha$.

\tit{Why Both Similarities Matter}
Existing metrics, primarily designed for raster images, struggle to capture the intrinsic characteristics of vector graphics. Image similarity in the projected space effectively captures local visual cues specific to vectorized representations, while textual similarity provides a global measure of semantic alignment.

For example, a text-to-SVG model may generate an SVG that differs structurally from the reference (\eg, a triangle with rounded edges instead of sharp corners) but preserves the intended concept. In this case, the semantic similarity $S_T$ would remain high, while the image similarity $S_I$ would drop. Conversely, a model might replicate the visual structure but misrepresent the intended meaning (\eg, generating a star when a flower was described), resulting in high $S_I$ but low $S_T$. By combining both scores, our metric remains robust to such discrepancies and achieves closer alignment with human judgment of SVG quality and semantic fidelity.

%% file: sections/experiments.tex
In this section, we assess the quality and human-alignment level of the proposed metric for text-to-SVG generation. Using a newly constructed dataset consisting of SVG-prompt pairs and human ratings, we explore various methodological combinations and benchmark our metric against established metrics (\lpips, \dinosimilarity, \fid, and \clipscore). Further, we also apply our metric in system-level correlation experiments, where we assess eight LLM-based generators under zero-shot conditions.

\subsection{The \dataset Dataset}
To assess the alignment of the proposed metric with the human judgment, we create and release the SVG Human-Evaluation dataset (\dataset)\footnote{The dataset will be publicly released upon paper acceptance.} -- a collection consisting of 333 SVG-prompt pairs, each associated with around eight generations coming from different models, evaluated with human scores. The collection of the original SVG images is initially obtained through web scraping and then manually assessed to ensure diversity in image complexity -- ranging from simple black icons to elaborate illustrations. The prompts accompanying the SVG images are instead automatically generated through a state-of-the-art Multimodal LLM, Idefics3 \cite{laurençon2024building}, and subsequently validated through human annotation.

\input{tables/user_Study}

To collect generated SVG images for the subsequent collection of human judgment, we select eight state-of-the-art LLMs, covering both small-scale models (\gemma, \llamas, \deepseeks, and \mistrals) and larger models (\chatgpt, \llamab, \mistralb, and \deepseekb). Generated outputs that failed to produce syntactically correct SVG code were excluded from the dataset, while syntactically valid yet visually blank outputs were retained to increase diversity and fairness of evaluation in system-level evaluations. In Table \ref{tab:user_study}, we report statistics on the generated results across the selected models. As can be observed, smaller LLMs typically struggle even with basic SVG image generation, irrespective of output quality, while larger models tend to provide syntactically correct SVG generations.

We then conducted a human evaluation study to collect human judgment annotations. The study involved 40 participants with diverse backgrounds, ranging from AI researchers to non-experts. Participants were asked to evaluate the correlation between input prompts and the corresponding generated SVG images on a scale from 1 (completely unrelated) to 5 (completely related). The study resulted in a total of 2,461 annotations, with an average of 57 evaluations per user. Visually blank generations were always evaluated with the lowest score.
Finally, the dataset is split into training and test subsets containing 2,000 and 461 samples, respectively. Table \ref{tab:user_study} also reports the average human rating obtained by each generator.

\subsection{Quantitative results}
\tit{Implementation details}
To assess the role of the visual backbone for computing the visual similarity score, we consider different image encoders, namely \dinovdfull, \maefull, \siglipfull, and \clipfull. SVG images are rasterized with a fixed white background, using the input resolutions employed by each encoder. Unless otherwise stated, we always consider the grid of activations coming from the last self-attention layer of the architecture, ignoring the $\texttt{[CLS]}$ token. For PCA computation, we retain the first $128$ largest eigenvectors. 
Additionally, to assess the role of the captioner when encoding the generated image, we consider three captioning models, namely \textit{Florence2-base} \cite{xiao2023florence}, \textit{Idefics3} \cite{laurençon2024building}, and \textit{Blip2} \cite{10.5555/3618408.3619222}.

To measure the correlation between the ratings of \ours and human evaluations, we employ the Spearman, Kendall, and Pearson correlation coefficients.

\tit{Ablation studies}
We first evaluate the impact of the selected image encoder and the choice of captioning model. In Table \ref{tab:cross_corr}, we present aggregated human correlation values, when employing feature vectors extracted from different visual encoders, and when verifying the captioner employed to generate the description of the generated image. In particular, we test by employing the $\texttt{[CLS]}$ token of the last self-attention layer, or by taking the average of the grid of features at the last self attention layer (excluding the  $\texttt{[CLS]}$) of CLIP, DINOv2 and SigLIP.

To represent the overall performance of each experimental combination, we aggregate the human correlation values across different values of $\alpha$ and $\beta$ (at evenly spaced values between 0 and 1.0) and by taking the average of the Spearman, Kendall, and Pearson correlation values. This yields a single aggregated score, which is formally computed as
\[
\frac{1}{C \times P} \sum_{\alpha} \sum_{\beta} S_{\alpha,\beta} + K_{\alpha,\beta} + P_{\alpha,\beta} ,
\]
where $C$ is 3 (the number of correlation coefficients) and $P$ is 11 (number of $\alpha$,$\beta$ combinations).

\input{tables/CC_ImageEnc_Captioner}

As shown in Table \ref{tab:cross_corr}, \blip~on average emerges as the most effective captioning model across all image encoders, while \siglip~proves to be the most effective image encoder. Interestingly, employing the average of the feature grid always provides better results than the $\texttt{[CLS]}$ token, underlining the effectiveness of considering spatial-aware features.

We then conduct further experiments to assess the impact of the different transformations applied to the visual feature vectors. Table \ref{tab:lhs_trans} reports the results obtained when applying a Generalized Mean Pooling \cite{8382272}~(with $p=1,2,4$) and PCA, with and without whitening. PCA is applied by transforming the average of the feature grid into a lower-dimensional subspace. When PCA is combined with whitening, data is centered by subtracting the mean, and each component is scaled to achieve unit variance, resulting in decorrelated features and alignment with the distribution of SVG images.

As can be noticed from Table \ref{tab:lhs_trans}, the best performance is achieved when combining PCA and whitening with \siglip~as visual feature extractor, and \blip~as captioner. Table \ref{tab:alphabeta_config} further illustrates the correlation with human evaluation across different values of $\alpha$ and $\beta$ on both training and test sets, underscoring the robust generalization capability of our proposed approach. Based on the results obtained from the training set, we select $\alpha = 0.6$  and $\beta = 0.4$ as the default values of our final metric configuration.

\input{tables/alpha-beta-config}

\tit{Comparison with the State-of-the-Art}
Finally, we compare our proposed metric against established alternatives for text-to-SVG generation -- namely \fid, \lpips, \dinosimilarity, and \clipscore, by performing experiments both at the system level and instance level. Table~\ref{tab:system_level_results} shows that, as per human evaluation scores, \chatgpt~is the best-performing model, followed by \llamab~and \mistralb. Notably, our metric produces a model ranking that closely mirrors human judgments, assigning the highest score to \chatgpt, followed by \mistralb~and \llamab. In contrast, existing older metrics such as \fid~and \lpips~either poorly correlate with human ratings or produce rankings that are not aligned with observed semantic relevance. Moreover, those metrics struggle to correctly rank generator models, which may be attributed to the absence of SVG-like data in their training distributions. It is also worth noting that lower FID scores indicate better performance, yet the average FID values observed in our experiments are significantly high, suggesting a poor fit of this metric to the SVG domain (\eg, icons, illustration, etc.).

\input{tables/sys_level}

This trend is also quantitatively confirmed in the system-level correlations reported in Table~\ref{tab:system_level_correlation_1}, where our metric achieves the highest correlation with human judgments across Spearman, Kendall, and Pearson scores. These results highlight the limitations of state-of-the-art automatic metrics in capturing the semantic alignment between SVG generated images and text prompts, and highlight the effectiveness of our proposed approach in producing evaluations that are more consistent with human perception.

In Table \ref{tab:system_level_correlation_2}, we report instance-level correlations in terms of Spearman, Kendall and Pearson correlation coefficients for four metrics (\lpips, \dinosimilarity, \clipscore, and \ours) in both reference-based and reference-free settings. For the purpose of this experiment and fairness of evaluation, we also show a reference-free version of \ours, where we remove the visual similarity portion of the metric (\ie, setting $\alpha = 0$). Reference-based \ours achieves the strongest agreement with human judgments. The reference-free variant of \ours trails \clipscore~in ranking metrics but surpasses it in Pearson correlation. Overall, our reference-free approach offers a competitive alternative to \clipscore. These results further underline that current spread-wise evaluation methods remain unreliable when applied to SVG-based content.

\input{tables/sys_level_corr}

\subsection{Qualitative results}
\begin{figure*}[tb]
    \centering
    \includegraphics[width=0.925\linewidth]{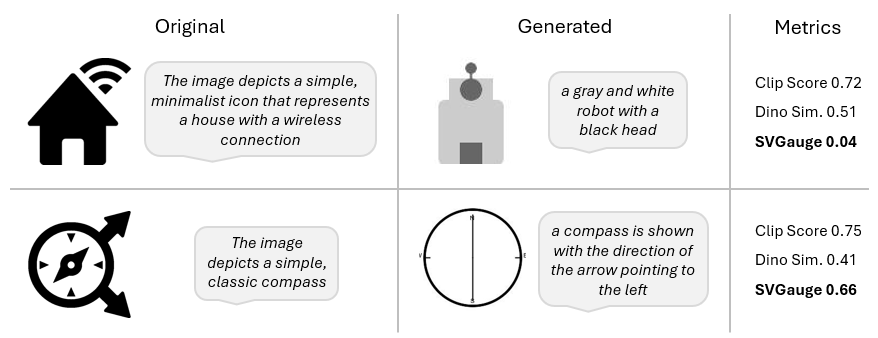}
    \caption{Sample qualitative results.
    }
    \label{fig:qualitatives}
    \vspace{-1em}
\end{figure*}

Finally, in Figure~\ref{fig:qualitatives} we showcase a qualitative comparison between \ours, \clipscore~and  \dinosimilarity. In the first row, although the generated SVG is visually and semantically unrelated to the reference image, \clipscore~and \dinosimilarity~still give relatively high scores. In contrast, \ours correctly assigns a low score, better aligned with human perception. On the other hand, in the second row, when the generated SVG accurately preserves the core semantics required by the prompt, despite the lower quality image, \ours correctly assigns a moderate score. These examples highlight that \ours features an improved sensitivity to semantic consistency, and demonstrates a significant ability to distinguish between meaningful and misleading generations in comparison with other existing metrics.

\vspace{0.1em}

%% file: tables/user_Study.tex
\begin{table}[tb]
  \caption{Dataset generation and user study statistics. Here, ``\% Generated'' indicates the percentage of times that the generator returned SVG code, ``\% Correct syntax'' the percentage of times the generated SVG code was syntactically correct, and ``\% Whites'' the percentage of times the generated SVG code resulted in a fully white image. We also report the average human score obtained over images generated by each model.}
  \vspace{-.5em}
  \label{tab:user_study}
  \centering
  \setlength{\tabcolsep}{.6em}
    \resizebox{0.99\linewidth}{!}{%
    \begin{tabular}{l cccc}
        \toprule
         & \textbf{\% Generated} & \textbf{\% Correct Syntax} & \textbf{\% Whites} & \textbf{Human Score} \\
        \midrule
        \deepseeks  & 93.4 & 100.0 & 21.9 & 1.21 \\
        \llamas     & 98.8 & 100.0 & 6.1 & 1.83 \\
        \gemma   & 92.8 & 100.0 & 64.6 & 1.90 \\
        \mistrals   & 93.4 & 100.0 & 1.6 & 1.80 \\
        \mistralb  & 100.0 & 96.4 & 0.3 & 2.49 \\
        \deepseekb & 98.5 & 98.8 & 5.3 & 2.23 \\
        \llamab    & 99.7 & 100.0 & 3.6 & 2.46 \\
        \chatgpt     & 100.0 & 100.0 & 0.0 & 3.37 \\
        \bottomrule
    \end{tabular}
    }
\vspace{-1em}
\end{table}

%% file: tables/CC_ImageEnc_Captioner.tex
    


\begin{table}[tb]
\begin{minipage}[c]{0.48\textwidth}
  \centering
  \caption{Average Human Correlation with different captioners and image encoders.}
\label{tab:cross_corr}
  \vspace{-.5em}
  \centering
  \setlength{\tabcolsep}{.1em}
    \resizebox{0.99\linewidth}{!}{%
    \begin{tabular}{c ccc }
        \toprule
        &\multicolumn{3}{c}{\textbf{CLS token}}\\
        \cmidrule(lr){2-4}
        & \textbf{\florence} & \textbf{\idefics} & \textbf{\blip} \\
        \midrule
   \clip   & 33.0 & 26.5 & 34.1 \\
    \dinovd & 36.2 & 29.7 & 37.1 \\ 
    
    \siglip & \textbf{37.5} & \textbf{30.5} & \textbf{38.8} \\
    \midrule
    &\multicolumn{3}{c}{\textbf{Mean Feature Grid}}\\
    \cmidrule(lr){2-4}
    \mae & 0.8 & -5.2 & -1.7 \\
    \clip & 32.5 & 23.2 & 33.9 \\
    \dinovd & 35.2 & 28.3 & 36.1 \\
    \siglip & \textbf{42.0} & \textbf{35.4} & \textbf{42.9} \\
    \bottomrule
    \end{tabular}
}
\end{minipage}
\hfill
\begin{minipage}[c]{0.48\textwidth}
\vspace{-1.5em}
    \centering
  \caption{Effects of transformation in the features space of each image encoder, when using \blip~as captioner.}
\label{tab:lhs_trans}
  \vspace{-.5em}
  \centering
  \setlength{\tabcolsep}{.3em}
    \resizebox{0.99\linewidth}{!}{%
    \begin{tabular}{c ccc cc}   
    \toprule
    & \multicolumn{3}{c}{\textbf{GeM Pooling}} & \multicolumn{2}{c}{\textbf{PCA}} \\ 
    \cmidrule(lr){2-4}\cmidrule(lr){5-6}
     & \textit{$p=1$} & \textit{$p=2$} & \textit{$p=4$} & w/o whit. & w/ whit. \\
     \midrule
    \mae & -1.7 & 30.4 & 31.6 & 32.1 & 32.1 \\
    \clip & 33.9 & 34.4 & 34.9 & 39.7 & 39.6 \\
    \dinovd & 36.1 & 36.1 & 36.2 & 38.7 & 38.5 \\
    \siglip & \textbf{42.9} & \textbf{36.3} & \textbf{36.2} & \textbf{44.1} & \textbf{44.3} \\
    \bottomrule
    \end{tabular}
    }
    \vspace{2.6em}
\end{minipage}
    \vspace{-1em}
\end{table}

%% file: tables/alpha-beta-config.tex
\begin{table}[tb]
  \caption{Correlation with human judgment for \ours with \siglip~and \blip~for different coefficients $\alpha$ and $\beta$. Correlation used are Spearman ($S\rho$), Kendall ($K\tau$) and Pearson ($Pr$). The higher the better for all correlation scores.}
  \vspace{-.5em}
  \label{tab:alphabeta_config}
  \centering
  \setlength{\tabcolsep}{2em}
    \resizebox{\linewidth}{!}{%
    \begin{tabular}{cc ccc ccc}
        \toprule
        \multicolumn{2}{c}{\textbf{Coefficients}}
          & \multicolumn{3}{c}{\textbf{Training set}} 
          & \multicolumn{3}{c}{\textbf{Test set}} \\
          
        \cmidrule(lr){1-2}\cmidrule(lr){3-5} \cmidrule(lr){6-8}
        $\alpha$
         & $\beta$
         & {\textbf{S$\rho$}} 
         & {\textbf{K$\tau$}} 
         & {\textbf{P$r$}} 
         & {\textbf{S$\rho$}} 
         & {\textbf{K$\tau$}} 
         & {\textbf{P$r$}} \\
        \midrule
        1.0 & 0.0 & 42.6 & 33.1 & 52.2 & 38.2 & 29.3 & 46.3 \\
        0.9 & 0.1 & 45.4 & 35.2 & 53.6 & 41.8 & 31.9 & 48.3 \\
        0.8 & 0.2 & 47.5 & 36.7 & 55.0 & 44.8 & 34.2 & 50.4 \\
        0.7 & 0.3 & 48.8 & 37.8 & 56.1 & 47.3 & 36.2 & 52.2 \\
        0.6 & 0.4 & \textbf{49.2} & \textbf{38.0} & 56.8 & \textbf{48.3} & \textbf{37.0}& 53.7 \\
        0.5 & 0.5 & 48.8 & 37.7 & \textbf{57.0} & \textbf{48.4} & \textbf{37.1 }& 54.6 \\
        0.4 & 0.6 & 47.3 & 36.4 & 56.5 & 47.2 & 36.1 & \textbf{54.7 }\\
        0.3 & 0.7 & 45.0 & 34.6 & 55.1 & 45.1 & 34.4 & 53.8 \\
        0.2 & 0.8 & 42.3 & 32.3 & 52.8 & 42.3 & 32.3 & 52.0 \\
        0.1 & 0.9 & 39.1 & 29.8 & 49.7 & 39.4 & 30.1 & 49.4 \\
        0.0 & 1.0 & 35.8 & 27.1 & 46.0 & 36.5 & 27.7 & 46.3 \\
        \midrule
        \rowcolor{blond}
        \multicolumn{2}{c}{\textbf{\raisebox{0em}[.85em][-1em]{Overall Mean}}}  
            & 44.7 & 34.4 & 53.7 & 43.6 & 33.3 & 51.0 \\
        \bottomrule
    \end{tabular}
    }
\vspace{-1em}
\end{table}

%% file: tables/sys_level.tex
\begin{table}[tb]
  \caption{System-level comparison between eight different generator models and different metrics.}
  \vspace{-.5em}
  \label{tab:system_level_results}
  \centering
  \setlength{\tabcolsep}{1em}
  \resizebox{0.99\linewidth}{!}{
\begin{tabular}{l ccccc >{\columncolor{blond}}c}
\toprule
 & \textbf{FID} & \textbf{LPIPS} & \textbf{\makecell{DINO\\Sim.}} & \textbf{\makecell{CLIP\\Score†}} & \textbf{Human} & \textbf{\ours} $\uparrow$ \\
\midrule
\deepseeks  & 656.43 & 0.61 &	0.32 &	0.65 &	1.28 &	\raisebox{0em}[1.1em][-1em]{\textbf{0.11}} \\
\mistrals & 497.95 &	0.56 &	0.40 &	0.70 &	1.86 &	\textbf{0.17} \\
\llamas &   541.41 &	0.59 &	0.38 &	0.66 &	1.66 &	\textbf{0.18} \\
\gemma &  800.39 &	0.59 &	0.29 &	0.62 &	1.66 &	\textbf{0.15}  \\
\mistralb & 485.95 &	0.56 &	0.40 &	0.71 &	2.31 &	\textbf{0.23} \\
\deepseekb & 502.27 &	0.58 &	0.42 &	0.72 &	2.23 &	\textbf{0.18}  \\
\llamab &   458.15 &	0.54 &	0.44 &	0.71 &	2.38 &	\textbf{0.22} \\
\chatgpt &   878.37 &	0.52 &	0.78 &	0.50 &	3.42 &	\textbf{0.25} \\
\bottomrule
\end{tabular}

}
\begin{flushleft}
  \scriptsize \vspace{-.5em} \hspace*{1em} † Reference free
\end{flushleft}
\vspace{-1em} 
\end{table}


%% file: tables/sys_level_corr.tex
\begin{table}[tb]
\centering
\begin{minipage}[c]{0.48\textwidth}
  \centering
  \caption{Comparison of system-level correlations. Correlation between the mean of human judgements for each generator model with other metrics.}
  \label{tab:system_level_correlation_1}
  \setlength{\tabcolsep}{.7em}
  \resizebox{\linewidth}{!}{
  \begin{tabular}{ccccc >{\columncolor{blond}}c}
    \toprule
    & FID & LPIPS & \makecell{DINO\\Sim.} & \makecell{CLIP\\Score†} & \textbf{\ours} \\

    \midrule
    \textbf{S$\rho$} & -7.14 & -76.2 & 76.2 & 80.9 & \raisebox{0em}[1.1em][-1em]{\textbf{91.0}} \\
    \textbf{K$\tau$} & -7.14 & -57.1 & 57.1 & 64.2 & \textbf{83.6} \\
    \textbf{P$r$} & 27.7 & -71.6 & 77.9 & 86.3 & \textbf{93.1} \\
    \bottomrule
  \end{tabular}
  }
  \begin{flushleft}
  \scriptsize \vspace{-.5em} \hspace*{.5pt} † Reference free
\end{flushleft}
\vspace{-2em}
\end{minipage}
\hfill
\begin{minipage}[c]{0.48\textwidth}
  \centering
  \caption{Instance-level correlation between \clipscore~and \ours both in reference-free and reference based versions.}
  \label{tab:system_level_correlation_2}
  \vspace{-0.3em}
  \setlength{\tabcolsep}{1.2em}
  \resizebox{\linewidth}{!}{
  \begin{tabular}{lccc}
    \toprule
             & \textbf{S$\rho$}
         & \textbf{K$\tau$}
         & \textbf{P$r$} \\
    \midrule
    LPIPS               & -14.0   & -10.7 & -16.5 \\
    Dino Sim.     & 34.6  & 26.5  & 39.3  \\
    \rowcolor{blond}
    \textbf{\raisebox{0em}[1em][-1em]{\ours Ref. based}}  & \textbf{48.4}  & \textbf{37.1}  & \textbf{53.7} \\
    \midrule
    {CLIP Score†}           & \textbf{37.8}  & \textbf{28.7}  & 40.6  \\
    \rowcolor{blond}
    \textbf{\raisebox{0em}[1em][-1em]{\ours Ref. free}}      & 36.5  & 27.7  & \textbf{46.3}  \\
    
    \bottomrule
  \end{tabular}
  }
  \vspace{-1em}
\end{minipage}
\end{table}

%% file: sections/conclusion.tex
We proposed the first evaluation framework specifically designed for text-to-SVG generation. By capturing both visual similarity and semantic alignment, our proposed metric \ours can provide quantitative evaluations with significant alignment with human perception. As a complementary contribution, we also developed the first dataset with prompt-SVG pairs annotated with human evaluations. We believe that our approach can serve as a foundation for future benchmarks and model evaluations in structured generative domains.